1# Performance and Energy Conservation of 3GPP IFOM Protocol for Dual Connectivity in Heterogeneous LTE-WLAN Network

Shubhada Gadgil, Shashi Ranjan and Abhay Karandikar: Department of Electrical Engineering,
IIT Bombay, Mumbai -400076, India {shubhadag, shashi.svk, karandi}@ee.iitb.ac.in**ABSTRACT**

For the 5th Generation (5G) networks, Third Generation Partnership Project (3GPP) is considering standardization of various solutions for traffic aggregation using licensed and unlicensed spectrum, to meet the rising data demands. IP Flow Mobility (IFOM) is a multi access connectivity solution/protocol standardized by the Internet Engineering Task force (IETF) and 3GPP in Release 10. It enables concurrent access for an User Equipment (UE) to Heterogeneous Networks (HetNets) such as Long Term Evolution (LTE) and IEEE 802.11 Wireless Local Area Network (WLAN). IFOM enabled UEs have multiple interfaces to connect to HetNets. They can have concurrent flows with different traffic types over these networks and can seamlessly switch the flows from one network to the other. In this paper, we focus on two objectives. First is to investigate the performance parameters e.g. throughput, latency, tunnelling overhead, packet loss, energy cost etc. of IFOM enabled UEs (IeUs) in HetNets of LTE and WLAN. We have proposed a novel mechanism to maximize the throughput of IeUs achieving a significant throughput gain with low latency for the IeUs. We have explored further and observed a throughput energy trade off for low data rate flows. To address this, we also propose a smart energy efficient and throughput optimization algorithm for the IeUs, resulting in a substantial reduction in energy cost, while maintaining the high throughput at lower latency and satisfying the Quality of Service (QoS) requirements of the IeUs.

*Keywords:*
IFOM, Dual connectivity, Heterogeneous networks, blocking probability, latency, throughput maximization, energy conservation, QoS.## 1. INTRODUCTION

Recently, there has been a significant explosion of cellular traffic. The Ericsson mobility report predicts the mobile data growth to skyrocket by 2022 with an average smart phone data usage of 66 EB per month [1]. Several innovative solutions are being proposed to meet the rising data demands. In a Heterogeneous Network (HetNet) scenario [2], i.e. a network with different Radio Access Technologies (RATs) e.g. IEEE 802.11 Wireless Local Area Network (WLAN) and Long Term Evolution (LTE), one solution is traffic steering, where selected Internet Protocol (IP) traffic is intelligently redirected to an alternate access network in the vicinity. The unlicensed 2.4 GHz and 5 GHz bands that the WLAN systems operate in, have been considered as important candidates to provide extra spectrum resources for cellular networks.

IP Flow Mobility (IFOM) [3] is one of the traffic steering /offload solutions standardized by Internet Engineering Task Force (IETF) [4] and Third Generation Partnership Project (3GPP) in Release 10 [5], that gains significance in this context. IFOM enables simultaneous connections to HetNets. IFOM enabled User Equipment (IeU)s have multiple interfaces; they can have concurrent flows with different traffic types over these network interfaces. They can also seamlessly switch these flows from one network to the other. IFOM is a promising solution that provides key advantages such as high bandwidth connections for the IeUs in the coverage of WLAN hotspots, enables the operators to manage the bandwidth and to provide different levels of service by applying different policies for UEs, tariffs and traffic types [6]. IFOM will play a prominent role in addressing the macro spectrum scarcity problem in HetNets and hence, we are inspired to pursue this vital 3GPP solution. For a seamless switchover with IFOM, all the flows are routed through a central junction, i.e. the Home Agent (HA) in the Packet Gateway (PGW) of the LTE core network. However, flow routing through a central junction point introduces some delay and involves signalling and tunnelling overhead, incurs energy cost and could affect the performance of the IeU. Hence, to investigate the performance metrics and enhance the user experience by optimum use of IFOM was an interesting research issue. Investigating the impact of concurrent flows in diverse network environments, interference, fading channel and varying load conditions are also issues that needed to be addressed. Although sizable literature describing the IFOM protocol exists [3, 5]–[7], few literature is available which investigates the benefits and implications of IFOM capability in HetNets. We have investigated these issues in our work.

Although, using WLAN technology for traffic steering has been widely considered as a solution for addressing the spectrum scarcity problem, the mobile operators have identified many challenges in this regard. There are still open issues that the industry needs to solve before the LTE-WLAN inter-working solution is feasible. One of the main challenge is to devise an intelligent network selection solution that allows the operators to steer traffic in a manner that maximizes user experience and also caters to the challenges at the boundaries between the 3GPP and non-3GPP Radio Access technologies (RATs) such as LTE and WLAN.

They are:

1. WLAN on availability: The IeUs may associate with WLAN on availability, without actually evaluating its capabilities in terms of signal strength compared to the LTE network; it may hence receive a weaker signal resulting in an unsatisfactory IeU experience.
2. Heavily loaded WLAN: In HetNets of LTE and WLAN, the IeU at the boundary of lightly loaded macro region may select a strong WLAN, which is heavily loaded. The result can be an unsatisfactory IeU experience.
3. Inadequate bandwidth**:** An IeU may associate with a WLAN that has a lower bandwidth in the back-haul, compared to the LTE network; it is currently associated with, resulting in a degraded signal for the IeU.
4. Battery drain and energy consideration**:** Association with a strong LTE network may result in more energy consumption compared to WLAN.

Presently, there are no standardized practices for network selection and traffic steering and hence, there is inconsistency in implementation of these solutions by the Operating Equipment Manufacturers (OEM). Also, traffic steering is in consideration with general UEs, where all traffic is offloaded and not for IFOM capable UEs with concurrent flows and dual connectivity features. Hence, the second objective in this research work is to explore solutions to steer traffic in a manner, so as to maximize the user experience, improve the QoS experience and provide an intelligent network steering behavior. We focus on enhancing the IeU performance metrics in terms of average per IeU throughput and reduced latency and energy cost in order to improve the QoS of the IeUs in HetNets. Currently, there's ongoing work to integrate multiple interfaces within the UE for the 5th Generation systems. However, 5G is yet to be standardised. Even the first version of 5G may take a year to be implemented. The first solution may not consider multiple access networks connectivity. WLAN is a specific interface and can be controlled better compared to a general interface. Hence, in this paper, we investigate IFOM with WLAN as an alternate network. Our algorithms are also relevant and significant for the latest 3GPP releases such as Release 13 LTE WLAN Aggregation (LWA).

We have proposed two algorithms; a novel throughput maximization algorithm to maximize the throughput of the IeUs and a smart energy efficient and throughput optimization algorithm to reduce their energy cost while maintaining high throughput.

The rest of the paper is organized as follows. Section 2 discusses related work and the contributions of the paper. Section 3 gives an overview of IFOM Protocol. In Section 4, we describe the system model, traffic models and energy models. We also discuss the classification of flows, signalling overhead for different traffic classes and our approach in selecting the appropriate access network for each flow. Section 5 elaborates our use cases. We also discuss the simulation scenarios for the use cases and present the results and inferences. We propose an algorithm to maximize the average throughput per IeU, present the throughput gain and evaluate the energy cost. We further propose a novel energy efficient and throughput optimizing algorithm for the IeUs and present the energy efficiency gain. Finally, we conclude the paper in Section 6.

## 2. RELATED WORK

IFOM enables selective and seamless switching of single or multiple flows associated with the same Packet Data Network (PDN) network. The operators can provide flow routing policies to the IeUs for selection of an appropriate 3GPP or a non 3GPP access network. This can be done either through Access Network Discovery and Selection Function (ANDSF) [8] or with static pre-configuration. In the following paragraphs, we discuss the literature and work related to IFOM.

In an overview of multi-access connectivity for mobile networking, the authors of [9] have discussed several solutions to enable multiple flows in a cellular access scenario including IFOM. They present a comparative study of these solutions in terms of location in the networking stack, support for mobility and flexibility in traffic scheduling. The authors discuss a subset of the solutions including network layer 3GPP IFOM, transport layer-Multipath Transmission Control Protocol (MPTCP) and application layer-Hyper Text Transfer Protocol (HTTP), which are deployable in today's mobile networks. Some solutions can perform negotiation of an available access network at connection establishment only e.g. Stream Control Transmission Protocol (SCTP). Some solutions like the application layer approaches can't re-negotiate the available access network at all, and thus, in case of mobility, they need to establish a new flow. While solutions like IFOM are able to re-negotiate the available access at all times (e.g. MPTCP, 3GPP IFOM and all Mobile IP variants). Seamless switch-over of flows is hence possible with IFOM.

The author in [7] presents an overview of the data offloading techniques in 3GPP Release 10 networks [3]. The author describes IFOM, Local IP Access (LIPA) and Selected IP Traffic Offload (SIPTO) offloading techniques for HetNets,

4and discusses their merits and shortcomings. LIPA/SIPTO offloads are defined for residential small cells operating in licensed spectrum. With the LIPA approach, an UE associated with a Home eNodeB (HeNB), can offload data to a local network connected to the same HeNB, without traversing the macro network. With the SIPTO approach, a part of the data traffic associated with HeNB or macro network can be transferred to a local network, thus relieving the load of the macro network. It is applicable in both femto and macro network use cases, but LIPA and SIPTO offloading does not help in enhancing spectrum capacity of the macro network. Moreover, the breakout in the context of LIPA/SIPTO is the location in the 3GPP architecture where the data offloading takes place [10]. The breakout could be at the private network (covering the LIPA and femto SIPTO cases) or at/above Radio Access Network (RAN) (covering macro SIPTO and femto SIPTO cases). Hence, switching of flows traversing from one network to other is not seamless and the offloading is network controlled.

The IFOM approach differs from the LIPA/SIPTO approach in its architecture, as in this case, an UE has multiple parallel interfaces, with which, it can connect to multiple RATs simultaneously. For e.g. an IeU can simultaneously connect to a 3GPP access and a non-3GPP WLAN access. It can exchange different IP flows belonging to the same Packet Data Network (PDN) between these access networks. IP flows belonging to the same or different applications can be switched seamlessly between these access networks. IP flow scheduling can be done flexibly on a per-flow basis. Switching of flows can be controlled by the IeU.

IFOM covers both licensed and unlicensed spectrum. It has been argued that IFOM would be helpful in relieving both radio and core network congestion, and can help in macro radio access capacity enhancement without much change on the network side. The only constraint is, it requires the support of Dual Stack Mobile Internet Protocol Version 6 (DSMIPv6) [11].

In [6], the authors present advantages and limitations of two standardized techniques that enable IP flow mobility: client-based and network-based IP flow mobility. It has been concluded that the network-based approach will be better for enhancing network capacity and tier service offerings in heterogeneous access networks at a reduced cost.

In [12], the authors investigate the potential benefits of flow based routing in multi homed environments. Their results show that for bandwidth intensive flows, the average capacity can be improved by a factor of three. However, their investigation is for non IeUs only. Throughput for concurrent flows on alternate accesses for IeUs needs to be evaluated. Along with it, energy consumption is also an important area which needs to be inspected.

The authors in [13] focus on energy efficiency of IFOM in the uplink case. They propose a scheme for offloading the uplink traffic of users that are within the coverage of WLAN Access Points (APs), with the assumption that the data requirements of UEs are known a-priori to the WLAN AP. However, it has been estimated that, more than 75% of the cellular traffic is in the downlink [14]. Hence, Energy efficiency in the downlink case needs to be evaluated. Also, blocking probability, packet loss, latency, throughput improvement along with energy conservation are the parameters that need to be addressed.

*2.1. Our Contributions:*

We have proposed a novel throughput maximization algorithm for the IeUs achieving a significant 30%-35% throughput gain, with negligible packet loss and marginal latency. We have observed a throughput energy trade off for low data rate application flows (FTP 256 Kbps, 512 Kbps). Hence, to overcome the throughput-energy trade-off, we have developed an opportunistic energy efficient and throughput optimization algorithm for the IeUs to intelligently switch the traffic flows between LTE and WLAN, in a way as to obtain the best throughput at a substantially reduced energy cost. We verify the same through our extensive $ns-3$ [15] simulations for different scenarios, for varying load, data rate and transmit power levels of LTE Base Station and WLAN Access Points (WAPs). Our analysis includes the impact of IFOM protocol on tunnelling overhead, packet loss, blocking probability, latency and energy consumption/conservation with the intelligent use of IFOM Protocol. Further, with our energy efficient algorithm, we are able to achieve a 20% reduction in the energy cost for the IeUs, while maintaining the high throughput at lower latency. We had to make significant additions to the $ns-3$ code to implement IFOM. We have carried out exhaustive simulations for all the presented scenarios. The results are presented in Section V.

## 3. DUAL CONNECTIVITY WITH IFOM PROTOCOL

An IeU (as illustrated in Figure 1) can handle data flows simultaneously on multiple interfaces with different access networks such as LTE, WLAN etc.. The IeU may have concurrent flows of different traffic types over these access networks, with an ability to switch a single IP flow selectively and seamlessly to a different radio access, while keeping rest of the ongoing connections intact. With IFOM capability, the IeU can hence get high throughput connections when in the coverage of WLANs. For client based IP mobility, the IeU is required to be involved in mobility management, running a specialized stack that can detect, signal and react to changes of point of attachment. IFOM uses Dual Stack Mobile Internet Protocol version 6 (DSMIPV6) [4] for this purpose. The IeU is involved in routing different traffic flows. As per the basic MIPv6 specifications [3], every UE has a Home Agent (HA), an entity located at its home network that provides the UE a permanent IP address called the Home Address (HoA).

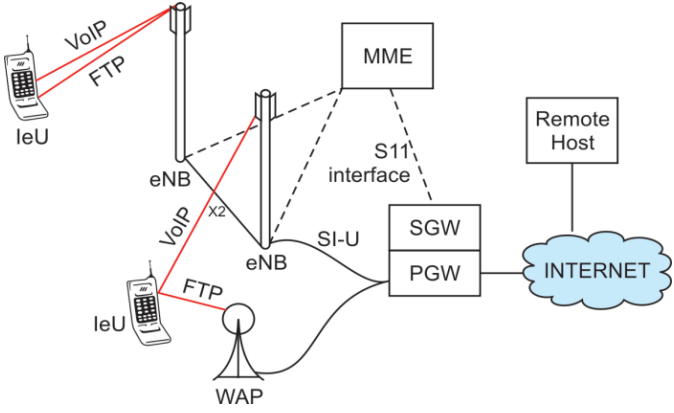

Fig. 1: IeU with concurrent flows over Macro and WLAN

As shown in the Figure 1, the HA is located in the PDN Gateway. While being away from the home network, it obtains a temporary IP address called Care of Address (CoA) from the visited network and informs the HA about it. A bidirectional tunnel is then set up between the HA and the UE, for the two way IP traffic flow. As seen in Figure 2, the UE can have several such CoAs with the foreign networks, but only one CoA, known as the primary CoA, is registered with the HA. It means that, the UE cannot use its multiple interfaces simultaneously. Therefore, to support IFOM, IETF has added three extensions to the existing MIPV6 specifications [6]:

1. Multiple Care of Address (CoA) registration-An IeU needs to be configured with multiple active IPv6 CoAs. The extensions to the Mobile IPv6 protocol standardized in [4, 16], support registration and use of multiple CoAs for a HA and create multiple binding cache entries. Binding cache as illustrated in Figure 2 is a cache maintained by the HA that has a different binding for each pair of CoA and HA. A new Binding Identification (BID) number is created for each binding the UE wants to create and is sent to the Binding Update (BU). A BU adds, modifies, refreshes or deletes a given binding.

2. Flow Bindings support-The IeU should be able to associate one or more IP flows with a specific CoA. The Flow Bindings extensions in MIPv6 and Network Mobility (NEMO) Basic Support standardized in [17] facilitate this.

3. Traffic Selectors - enable identification of traffic flows. IP flows can be identified based on a subset of parameters such as source/destination address, source/destination port, traffic class, flow label, etc.. Based on the type of flow, the IeU can instruct the HA to route inbound and outbound packets of specific flows with specific CoAs.

The procedures for IFOM association and disassociation are defined in 3GPP [5] but, it is desirable to verify the practical aspects of these procedures regarding how they affect the behaviour of the system. To do that, we need to model the LTE system accordingly. We have implemented the same in our ns − 3 based simulator. We describe in the following

Fig. 2: Binding Cache of HA with Multiple CoA Registrations

section, our system model, energy and traffic models for the study of IFOM.

## 4. SYSTEM MODEL FOR THE STUDY OF IFOM PROTOCOL

We consider the 3GPP specified urban macro deployment of 21 sectors, 7 hexagonal macro cells having an inter-site distance of 500 m [18]. Macro BSs with 3 sectored directional parabolic dish antennae are deployed at the center of each cell. UEs are dropped randomly, following uniform distribution in each sector. Hotspot regions (with higher UE densities) are positioned with their centers at a distance of 120 m from LTE in each sector. The WLAN APs are deployed one each, in each sector at the center of the hot-spots. We assume that the WLANs are deployed by the mobile operators. Hence, the locations of the WLAN APs are apriori known. We drop a single UE in the rest of the sectors of the surrounding cells to create interference to the UEs in the central cell. Shadow fading is modelled as a log-normal random variable with mean zero and standard deviation of 8 dB for both LTE and WLAN [18]. The simulation parameters indicated in Table 1 for the Macro BSs, WLAN Access Points (AP) and UEs as proposed by 3GPP [18] and [19] are adopted for our simulations. The system is modelled using the network simulator ns − 3. We had to make significant additions to ns −3 code and additions to ns − 3 modules for implementation and study of IFOM.

### 4.1 Traffic Models for IeU Nodes:

We consider VoIP, video and FTP traffic models [19] for our simulations. We set up a bearer that provides a dedicated tunnel with Guaranteed Bit Rate (GBR) for VoIP and video traffic with Quality of Service Class Identifier (QCI) set to 1 and 2 for VoIP and video, respectively (as per the guidelines of [20]). QCI is the QoS information associated with the EPS bearer. A QCI of 1 and 2 sets the Packet Delay Budget (PDB) and Packet Error Rate (PER) for VoIP and video as indicated in Table 2 and the priority level to 2 and 4, where a lower number indicates higher priority. We choose the full data rate of 12.2 Kbps for VoIP to capture the worst case scenario. The FTP traffic is over the default bearer.

**Table 1: Simulation parameters**

| Parameter | Value |
| --- | --- |
| Antenna height of eNB, UE and WAP | 32 m, 1.5 m and 2.5 m |
| Tx power of eNB and UE | 46 dBm and 23 dBm |
| Noise Margin of eNB and UE | 9 dB and 5 dB |
| eNB to UE minimum distance | 35 m |
| UE antenna pattern | Isotropic |
| Operating Bandwidth per DL | 10 MHz |
| WLAN Standard | IEEE 802.11g, 54 Mbps |
| Range of WAP | 60 m |
| WAP Tx Power | 23 dBm |
| Noise Margin of WAP | 4 dB |
| Path Loss for LTE Network | 128.1 + 37.6 log 10(R) |
| Path Loss for WLAN | 140.7 + 36.7 log 10(R) |
|  | R (distance in Kms) |

**Table 2: Traffic Models**

| VoIP Traffic | Parameters |
|---|---|
| Data Rate | 12.2 Kbps |
| Packet Size | 33 bytes |
| Packet Delay Budget | 100 ms |
| Packet Error Rate | $10^{-2}$ |
| **Video Traffic** | **Parameters** |
| Data Rate | 1.5 Mbps |
| Packet Size | 250 Bytes |
| Packet Delay Budget | 150 ms |
| Packet Error Rate | $10^{-3}$ |
| **FTP Traffic** | **Parameters** |
| Data Rate | 256, 512 and 1024 Kbps |
| Packet Size | 1024 bytes |

We consider a scenario where each IeU has two concurrent flows, VoIP/video and FTP in the downlink.

In our algorithms, we apply the following traffic flow rules to the IeUs − 1) Real time flows such as VoIP and video, that require special Quality of Service treatment from the 3GPP network, stay with the macro network. 2) Non-real time flows such as FTP do not need any special treatment from the 3GPP network. For energy efficiency, cost effectiveness and lower latency, whenever WLAN access is available, we maintain maximum flows with WLAN and rest with LTE.

Three categories of UEs are considered, namely LTE-only IeUs, LTE-WLAN IeUs and Background UEs. LTE-only IeUs have both VoIP/video and FTP flows associated with LTE only. LTE-WLAN IeUs are the IeUs in the coverage of WLAN and have VoIP/video flows associated with LTE and FTP flows associated with WLAN. The background UEs are ordinary UEs with a single FTP flow in the downlink. When the number of IeUs is low, the background UEs are dropped in the central macro cell to contribute to the network load. They associate their FTP flows with LTE and offload them to WLAN APs, whenever in the coverage of WLAN.

### 4.2 Energy Models for LTE BSs and WLAN APs

The energy models adopted in our simulator are discussed below.

*4.2.1 WLAN AP Radio Energy Model:* The WLAN energy model has a current breakup as shown in Table 3. The model is already a part of WiFi ns − 3 module [21]. The WiFi radio has four different states – Transmit (TX), Receive (RX), IDLE and SLEEP. Table 3 lists the typical value of current breakup of this module.

**Table 3: WLAN AP radio current breakup**

| Parameter | Symbol | Value |
|---|---|---|
| Current in TX state | $I_{tx}$ | 380 mA |
| Current in RX mode | $I_{rx}$ | 313 mA |
| Current in Sleep mode | $I_{sleep}$ | 33 mA |
| Current in idle mode | $I_{idle}$ | 273 mA |

*4.2.2 LTE Energy Model:* We adopt a linear energy consumption model in the downlink for LTE Base Station and the IeUs [22]–[24]. We denote $P_{idle}$ as the idle power or the static power consumption, $P_{rx}$ as the receiver power consumption and $P_{tx}$ as the transmit power consumption. Also, let $t_{idle}$, $t_{tx}$ and $t_{rx}$ be the time spent in idle state, transmit state and receive state, respectively. The energy consumption in the idle state comprising static power is,

$$E_{idle} = P_{idle} \, t_{idle}. \quad (1)$$

The energy consumption in the transmit state is,

$$E_{tx} = (P_{idle} + \alpha P_{tx}) \, t_{tx}. \quad (2)$$

The energy consumption in the receive state is,

$$E_{rx} = (P_{idle} + P_{rx}) \, t_{rx}. \quad (3)$$

The total energy consumption of the radio device is,

$$E_T = E_{idle} + E_{rx} + E_{tx}. \quad (4)$$

The typical values of these parameters are listed in Table 4.

**Table 4: LTE Base Station energy model parameters**

| Parameter | Symbol | Value |
|---|---|---|
| Power in idle mode | $P_{idle}$ | 90 W |
| Transmit Power | $P_{tx}$ | 40 W |
| Linear coefficient | $\alpha$ | 4.27 |

### 5. SCENARIOS AND INFERENCES

We identify two use case scenarios with five different setups. In Scenario 1, we simulate IeU nodes and investigate the benefits of IFOM in HetNets. The objective as discussed in part-1 of the introduction is to make a thorough analysis of the user perceived performance of IFOM in terms of throughput, latency, energy cost, tunnelling overhead, blocking probability and packet loss for IFOM conforming UEs in the downlink. With ns − 3 simulations, we investigate the same for different configurations such as changing load, diverse networks with fading channel and interference, varying data rates, different traffic types and transmit levels of LTE base station and WLAN APs.

### 5.1 Scenario 1: Throughput and Energy Cost for Light, Medium and Heavy Load Conditions:

We introduce two IeUs, one LTE-only IeU and one LTE-WLAN IeU, each with two flows in the downlink -FTP flow with a data rate of 256 Kbps and VoIP flow with data rate of 12.2 Kbps in each sector of the central macro cell. We also introduce background UEs with a data rate of 512 Kbps and packet size of 1024 bytes. We drop the background UEs in the central cell uniformly and randomly upto the saturation levels of LTE eNBs and WLAN APs.

In our previous work [25] for a similar scenario, we have investigated throughput and latency for the IeUs in the worst case and best case situations. We have observed a guaranteed VoIP throughput of 12 Kbps for both the IeUs at varying background load. We have also observed the VoIP latency, to be below the standard tolerable range (50 ms as specified in [19]). However, the VoIP latency experienced by LTE-only IeU was slightly greater than LTE-WLAN IeU. This may be because of a single queue at the LTE BS as both flows are through LTE only. Similarly, we have observed that the FTP throughput for both the IeUs was high, even for worst case scenario, with a major drop only when the background traffic was high. The FTP throughput observed was above the minimum outage threshold of 128 Kbps as per the specifications in [19].

In this paper, we inspect the energy, tunnelling overhead and packet loss aspects under LTE-WLAN network.

*5.1.1 Tunnelling Overhead, Latency Overhead and Energy Cost:* With IFOM, all the flows associated with the non-3GPP access networks are routed through the Home Agent in the Packet Gateway of the macro (3GPP) network. IFOM uses the IPv6 routing header type 2 to route the flows associated with WLAN through the macro network. This introduces a tunnelling overhead between 24 bytes per packet [26] up to 68 bytes per packet if Advanced Encryption System (AES) with cipher is adopted [27]. In the current scenario, considering 68 bytes header length, $H_L$, for a packet size S of 1024 bytes, the tunnelling overhead, $T_H$ is $H_L/S * 100$ i.e. 6.7% of the packet size.

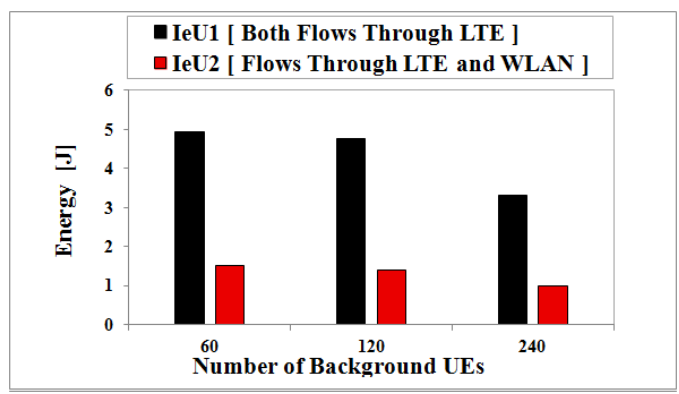

Fig. 3: Energy cost for varying system load (LTE and WLAN) for FTP traffic of IeUs

As LTE-only IeUs and LTE-WLAN IeUs associate their FTP flows with LTE and WLAN respectively, the energy cost is accordingly depicted in Figure 3. We can see that the energy cost for WLAN association is significantly lower than the cost for LTE association. The reason being that, the transmit power requirement for WLAN is very low compared to 3GPP macro cells, due to the short distance between transmitter and receiver.

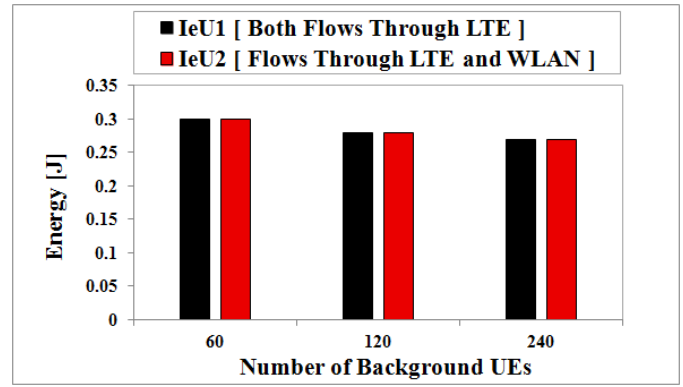

Fig. 4: Energy cost for varying system load (LTE and WLAN) for VoIP traffic of IeUs

The energy cost for VoIP flows are low (in the range of 0.25 J to 0.3 J) as noted from Figure 4. Also, the energy graphs for both LTE-only and LTE-WLAN IeUs coincide as VoIP flows for both the IeUs are through LTE only.

Combining the previous work and the above result, we can conclude that the IeUs with concurrent VoIP and FTP flows have high throughput, low latency and marginal tunnelling overhead. The IeUs experience lower latency and reduced energy cost in the coverage of WLANs.

*5.2 Scenario 2: Throughput Maximization and Energy Efficiency:*

As discussed in the introduction, the second objective in this research work is to explore solutions to steer traffic in a manner, so as to maximize the user experience, improve the QoS experience and provide an intelligent network steering behaviour. We focus on enhancing the IeU performance metrics in terms of average per IeU throughput and reduced latency and energy cost in order to improve the QoS of the IeUs in HetNets.

To meet this objective, we first conduct $ns-3$ simulations to characterize the LTE-WLAN heterogeneous network load capacity with all LTE-WLAN and LTE-only IeUs and investigate traffic aggregation and energy conservation with IFOM. We develop a throughput maximization algorithm for the IeUs, that intelligently switches the FTP traffic flows between LTE and WLAN based on certain decision metrics, in order to maximize the per IeU throughput. We also investigate the packet loss and blocking probability of the application flows. In scenario-1, we have evaluated the performance of IeUs with a background load of general UEs for energy and similarly in our previous work for throughput. Here, we inspect traffic aggregation and energy conservation with IFOM with all IFOM enabled LTE-only IeUs in Setup-1 and all LTE-WLAN IeUs in Setup-2. In Setup-3, we propose two algorithms: 1) Throughput maximization algorithm and 2) An energy efficient and throughput optimization algorithm to reduce the energy cost of the IeUs while maintaining high throughput.

***5.2.1 Setup 1: Network load characterization, Energy Cost with all LTE-Only IeUs:*** We drop all LTE-only IeUs in the central macro cell, having concurrent video and FTP flows in the downlink. All the video and FTP flows of the IeUs are associated with LTE only. We increase the network load by increasing the LTE-only IeUs gradually, until the average throughput per IeU drops below 65% of the maximum throughput. We chose this throughput threshold to ensure a better QoS for the IeUs. Also, we have observed through our simulation results that when the throughput falls below 60% of the maximum achievable throughput, the video flows are not scheduled, due to lack of Physical Resource Blocks (PRBs).

At this point, we define 'Light', 'Medium' and 'Heavy' load of the system as when number of IeUs is 36, 54 and 81, respectively. We plot the average per IeU throughput for Light, Medium and Heavy load for all setups in Figure 5 and the total energy cost for the same in Figure 6. The Cumulative Distribution Functions (CDFs) of the average throughput per IeUs for all setups are plotted in Figure 7 and the CDFs of the total energy cost of the IeUs in Figure 8, respectively. We make a proportionate study of the throughput, energy cost, tunneling overhead and packet delay for the same.

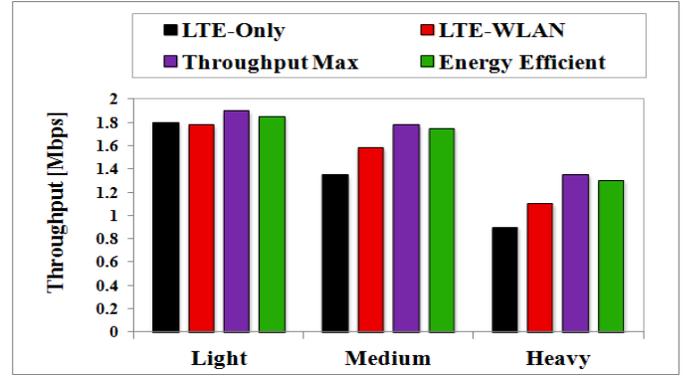

**Fig. 5: Average Throughput (video and FTP) per IeU**

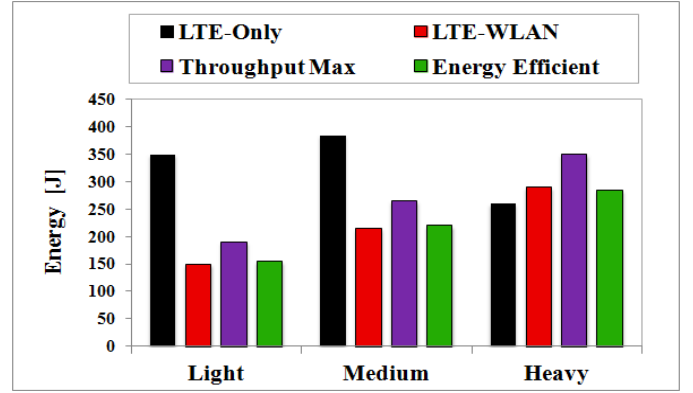

**Fig. 6: Total Energy Cost (video and FTP) of all IeUs for varying load**

At light load, as seen in Figure 5, we observe high throughput for LTE-only IeUs. The CDF in Figure 7a illustrates that 70% of the LTE-only IeUs experience high throughput at light load, whereas approximately 30% drop in the average throughput per IeU is observed at medium load. At heavy load, LTE is closer to saturation. As seen in Figure 5 and CDF in Figure 7c, we observe a significant 50% drop in the average (video and FTP) throughput per IeU. Video and FTP flows are dropped for as many as 30%-40% of the LTE-only IeUs. This is due to lack of PRBs and increased interference.

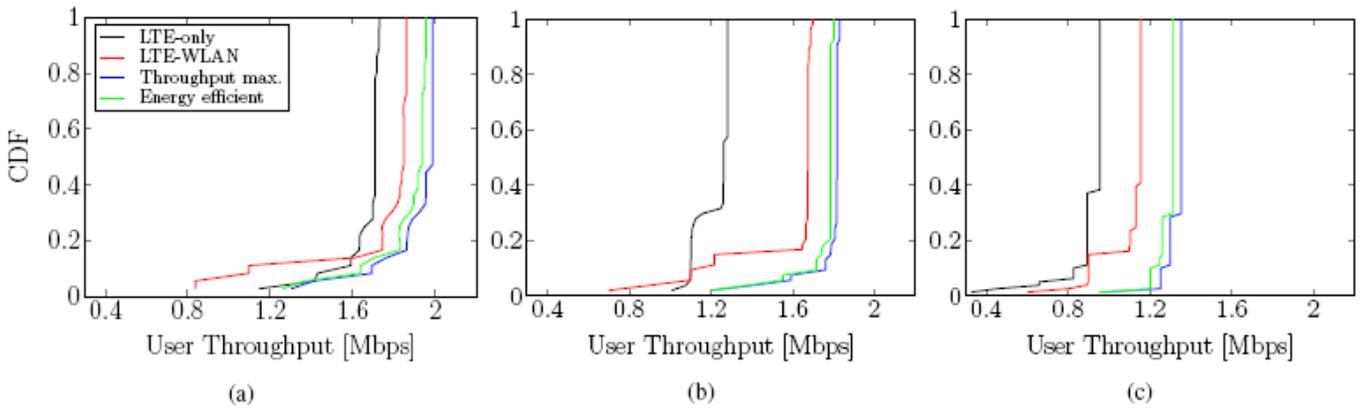

**Fig. 7: Throughput [Video + FTP] per IeU for (a) 'Light', (b) 'Medium' and (c) 'Heavy' load conditions**

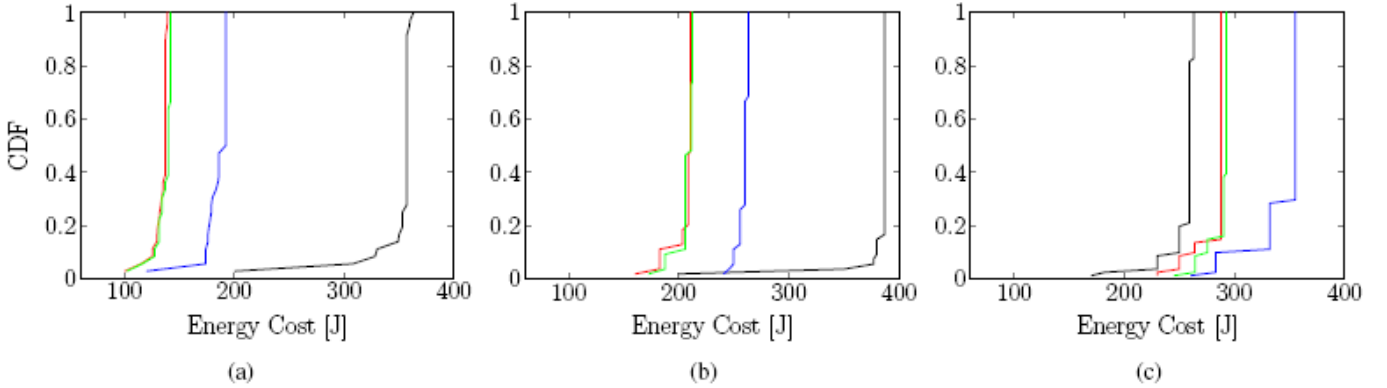

Fig. 8: Total energy cost [Video + FTP] per IeU for (a) 'Light', (b) 'Medium' and (c) 'Heavy' load conditions

*5.2.2 Tunnelling Overhead, Latency Overhead and Energy Cost:* In the current scenario, both the flows are through LTE only. The flows are not routed through the non 3GPP access; hence, there is no routing header and tunnelling overhead. The video latency is low in the range of 2.7 to 65 ms and below the 100 ms tolerable range as specified in 3GPP TR [18] and [19]. We now investigate the energy cost. As we increase the load from light to medium, the total energy cost of LTE-only IeUs increases as seen in Figure 6. As the load increases further, LTE becomes closer to saturation. We observe that the energy cost for the high data rate video flows associated with LTE is considerably low compared to lower data rate FTP flows associated with LTE. We observe that, as the data rate increases, the energy consumption increases only slightly. As the data rate increases, the number of transferred bits increases faster than the network energy consumption. The reason is that the power model is associated with a fixed static power consumption cost at zero RF output power and when the traffic increases, this cost is shared over a larger number of bits, which results into the energy per bit decrease.

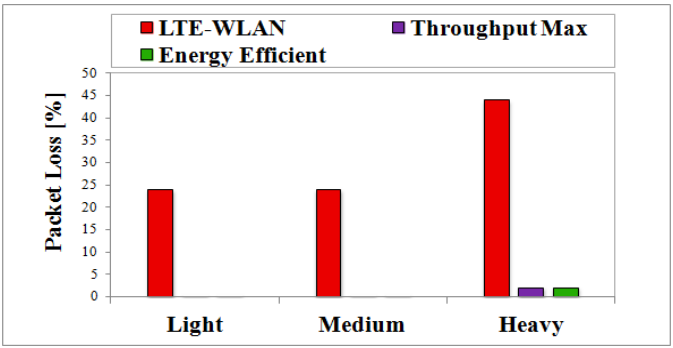

Fig. 9: Packet loss with and without algorithms

Also, as per [22] for LTE, the decoder power consumption does not scale linearly with the increased data rate. Increasing the data rate by a factor 10 only increases the power consumption about 5 percent. This implies that it is more energy-efficient to run at high data rates. At heavy load, we observe more number of video flows with stringent QoS requirement (discussed in Section IV-A) supported by LTE over the FTP flows. Hence, the FTP throughput drops drastically. The energy cost for video flows being lower, this results in a drop in the total energy cost, as seen in Figures 5 and 8c.

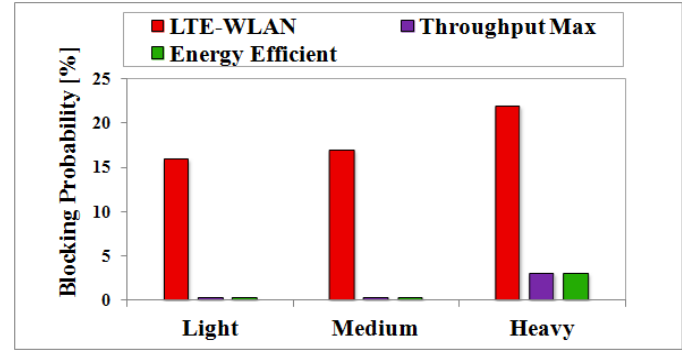

Fig. 10: Blocking probability with and without algorithms

*5.2.3 Setup 2: Network load characterization, Energy Conservation and Traffic Aggregation with IFOM (all LTE-WLAN IeUs-Dual Connectivity):* We now drop all LTE-WLAN IeUs having concurrent video and FTP flows in the downlink, in the central macro cell. All video flows of the IeUs are associated with the LTE network and all FTP flows are associated with WLAN.

At light load, since LTE and WLAN are under loaded, the throughput is good for both LTE-only and LTE-WLAN cases as seen in Figure 5. However, at medium and heavy load, the throughput observed is higher for the LTE-WLAN IeUs compared to the LTE-only IeUs, since the load is now distributed between LTE and WLAN. The throughput increase is about 20-22% at medium and heavy load, compared to LTE-only case. As indicated by the CDF in Figure 7c, the throughput (video and FTP) per IeU even at heavy load is high for nearly 70% of the IeUs. However, based on the position of the IeUs and the received signal strength, the FTP throughput of certain LTE-WLAN IeUs is very low, as they suffer high packet loss. The packet loss as seen in Figure 9 and blocking probability as illustrated in Figure 10 increases at heavy load. The video throughput with LTE association remains fairly good. However, some IeUs have very low FTP throughput. This may be because some IeUs are at the edge and hence receive weaker signal. We observe that for certain LTE-WLAN IeUs, FTP flows are not scheduled at all, as WLAN saturates; hence, they have only a single video flow that is associated with LTE.

*5.2.4 Tunnelling Overhead, Packet Delay overhead and Energy Cost:* Since, all FTP flows are associated with WLAN, the routing header overhead is 6.7% per packet for all FTP flows. As seen in Figure 6, the average energy cost per dual connected LTE-WLAN IeU is observed to be very low compared to the LTE-only IeUs. At light to medium load, the energy cost of LTE-WLAN IeUs is nearly 60% lower compared to the energy cost of LTE-only IeUs. We calculate the energy efficiency gain as $E = (E_2 - E_1)/E_2$ where $E_1$ and $E_2$ are the average energy costs for LTE-WLAN IeUs and the LTE-only IeUs, respectively. We find that the percentage energy efficiency gain $E$ varies between 20% to 60% with varying load.

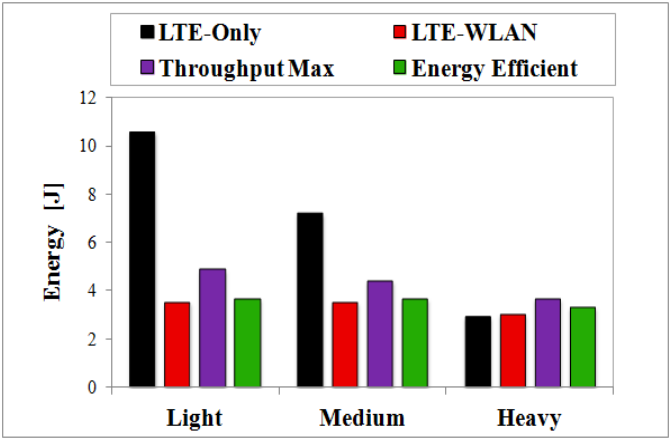

Fig. 11: Average energy cost per IeU (video and FTP (512 Kbps))

*5.2.5 Setup 3: Throughput Maximization and Energy Efficient Algorithms:* In this setup, we propose our throughput maximization and energy efficient algorithms. These algorithms are running on LTE eNBs and WLAN APs and for uplink can run on IeUs that are in dual region.

*5.2.6 Proposed Throughput Maximization Algorithm:* With an objective to maximize the throughput, minimize packet loss and blocking probability, improve the Quality of Experience (QoE) of the LTE-WLAN IeUs, we propose a novel throughput maximization algorithm as shown by the flowchart in Figure 12.

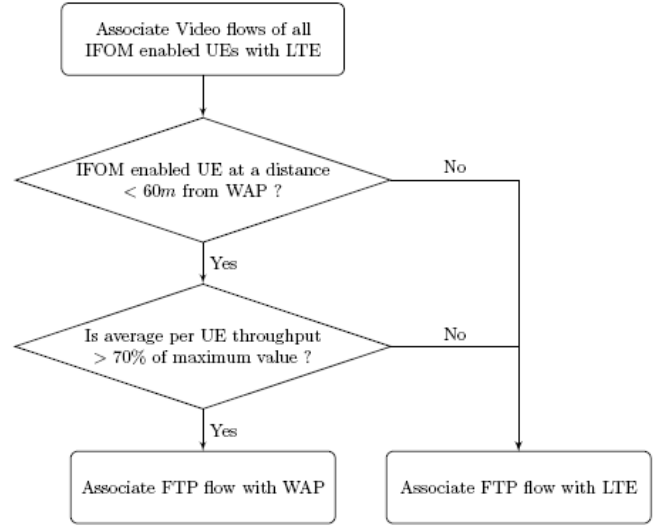

Fig. 12: Throughput maximization algorithm

The algorithm works as follows,
1. Retain all the video flows through LTE only. This is to ensure a good QoE for the IeUs.
2. Associate the FTP flows with WLAN, provided, the distance of the IeU from the AP is within 60 m and the average FTP throughput per IeU is above 70% of the maximum achievable throughput, otherwise, associate the flow with the LTE network. The distance information can be retrieved by the IeU or eNB on demand via the Access Network Discovery and Selection Function (ANDSF) server [8].

There may be few IeUs that have premium subscription. Both of their flows are associated with LTE network to guarantee QoS. The above algorithms are run for non-premium IeUs.

*5.2.7 Proposed Energy Efficient and Throughput Optimization Algorithm:* We propose a smart energy efficient and throughput optimization algorithm with an objective to minimize the average energy cost per dual connected LTE-WLAN IeU, while maintaining the high throughput with marginal blocking probability and packet loss. This algorithm is depicted in the flowchart in Figure 13.

The algorithm works as follows,
1. Video flows are retained on LTE to ensure a good QoE for the IeUs.
2. FTP flows are associated with LTE or WLAN based on the following policy – at light load, associate all FTP flows with WLAN, provided they are within a distance of 60 m from the AP. This is done as the cost to associate low data rate flows with WLAN is very low compared to the cost to associate them with LTE as discussed in Section V-B1. The distance threshold is to ensure that the quality of the received signal strength is good. The extensions to IEEE 802.11u [28] include additional fields to the APs beacon that provide the Channel utilization percentage (60% utilization indicates 60% of channel loaded) information. The IeU already has the beacon frame information and is updated in every beacon interval or can obtain it on demand with the Access Network Discovery and Query Protocol (ANQP). This information can assist the IeU to make an informed access network association decision.

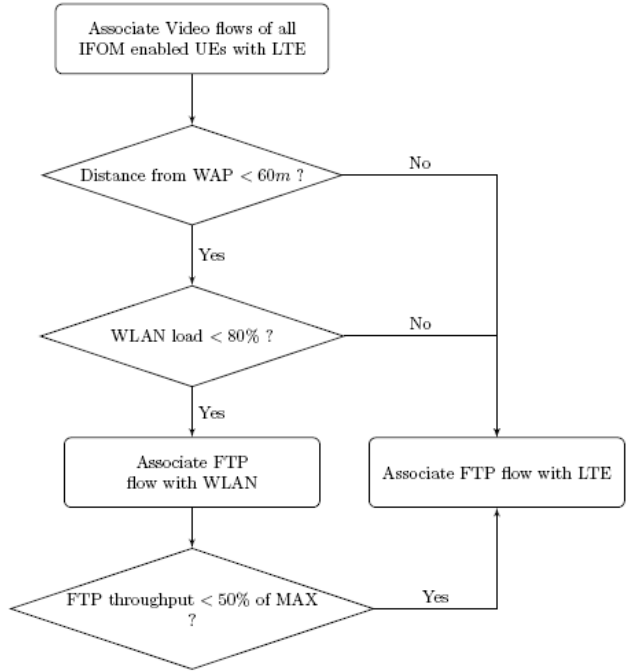

**Fig. 13: Energy efficient algorithm with throughput optimization**

3. If WLAN is loaded beyond 80%, the FTP flows are switched to LTE. This is done at medium to heavy load, since as the FTP throughput through WLAN deteriorates; the percentage packet loss increases along with the blocking probability.
4. Below 80% load, if the throughput through WLAN is greater than 50% of maximum throughput, the flows are retained on WLAN, else switched to the network providing better throughput.

From Figure 5, we observe that at light and medium load, both the throughput maximization and energy efficient algorithms work exceptionally well. The highest throughput is with the throughput maximization algorithm. It can also be observed that, at heavy load, the throughput gain increases with both the algorithms. This behaviour is due to the fact that at heavy load, the average FTP throughput for certain LTE-WLAN IeUs deteriorates, based on the position of the IeU and the received signal strength. In such situations, the algorithms effectively perform load balancing between LTE and WLAN. This keeps the average FTP throughput per IeU high contributing to an improved QoE for the IeU. From the CDF plots in Figures 7a and 7b, it can be seen that about 90% of the IeUs experience high throughput by both algorithms in comparison to without algorithms. With both the algorithms, the video flows are routed through LTE. Hence, we observe that the video throughput stays moderately stable upto 70% of maximum throughput.

*5.2.8 Tunnelling Overhead, Packet Delay overhead and Energy Cost:* From the plot of average per IeU throughput in Figure 5 and corresponding total energy cost in Figure 6, we observe the highest throughput with the throughput maximization algorithm, but also an increase in the energy cost. For nearly the same throughput, we observe a 20% reduction in the energy cost with the energy efficient algorithm. The video latency overhead remains nearly the same in both algorithms as the video flows are routed through LTE only. We check the latency for video flows and observe it to be lower for both the algorithms in the range of 2.4 ms to 48 ms below the standard tolerable limit of 100 ms as specified in [19]. For both the algorithms, the packet loss as seen in Figure 9 is zero for low to average load and marginal at heavy load. The blocking probability as shown in Figure 10 is also negligible. We observe that, with throughput maximization algorithm, the throughput gain for LTE-WLAN IeUs is significantly high with negligible packet loss and blocking probability, but has a trade-off with energy cost for low data rate flows. For nearly the same throughput gain and maintaining the negligible blocking probability and packet loss, we achieve a significant 20% reduction in energy cost with the energy efficient algorithm and 60% over the LTE only case.

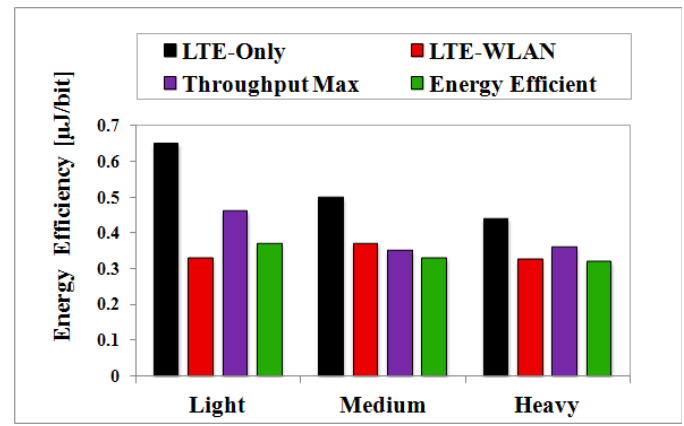

**Fig. 14: Energy cost for IeUs with (video and high data rate FTP (1 Mbps))**

We repeat the simulations with both algorithms for IeUs having concurrent video and high data rate FTP flows. We maintain the same data rate for video but a higher data rate of 1 Mbps for FTP. Figure 14 shows the total energy cost for the IeUs for varying load. We observe a comparatively lower energy cost for associating high data rate FTP flows with LTE compared to earlier low data rate (256, 512 Kbps) FTP flows. The FTP throughput has increased by 50% in the simulated scenario, and therefore the IeU can receive the data faster with marginal increase in energy consumption. Hence, the energy cost per bit is low. The difference between the energy cost of associating FTP flow with LTE to associating with WLAN reduces. Hence, the energy cost of IeU for concurrent video and FTP flows is very low. Energy Efficiency gain is between $5 - 20\%$ for varying load.

Further, to analyze the performance of our algorithms, we use the metric, User Satisfaction ($U_{sat}$) as the sum total of the throughput of all the IeUs divided by the product of the maximum throughput ($T_{max}$) of the IeUs and the total number of IeUs (K) in the macro cell region [29]. This metric expresses the relative throughput of an IeU compared to the throughput of the IeUs in the same cell and it indicates how close the IeU's throughput is to the maximum throughput. Let $T_{put}$ (k) be the throughput of the kth IeU.

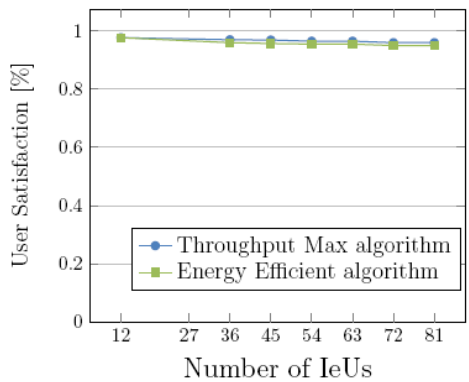

Fig. 15: User Satisfaction Plot for Algorithms

We check the $U_{sat}$ metric.

$$U_{sat} = \frac{\sum_k T_{put}(k)}{T_{max} * K} \quad (5)$$

$U_{sat}$ ranges between 0 and 1. When $U_{sat}$ approaches 1, all IeUs in the macro cell region experience nearly same throughput and when $U_{sat}$ approaches 0, there are big variations in the throughput achieved by the IeUs in the macro cell region. $U_{sat}$ has been selected as the performance metric since it leads to a fairer overall network performance.

As seen in Figure 15, we observe nearly similar throughput for all the IeUs with both the algorithms.

## 6. CONCLUSION

The 5G era will witness emergence of tight interworking with WLAN in a big way. Seamless and proficient interworking protocols such as IFOM will form the backbone of the next generation wireless cellular networks. Traffic steering mechanisms such as IFOM will gain significance with ever increasing data usage and huge number of home and WLAN networks spanning the entire geographical area.

In this paper, we have proposed a novel energy efficient and throughput maximization scheme for the IFOM enabled UE devices to substantially reduce their energy cost while enhancing their throughput. Our proposed throughput maximization algorithm results in a high throughput gain for the IeUs at marginal latency overhead, with negligible blocking probability and packet loss. This algorithm can be statically pre-configured in the IeU devices by network operators. The proposed smart Energy efficient and throughput optimization algorithm results in a significant 20% reduction in energy cost, while maintaining the high throughput gain at a similar latency overhead with negligible packet loss and blocking probability. The algorithm can be used in the power save mode. These algorithms are also relevant and significant for the latest 3GPP releases such as Release 13, LTE WLAN Aggregation (LWA).

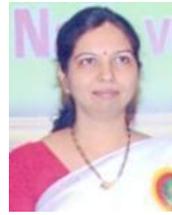

Shubhada Gadgil: Prof. Shubhada Rajesh Gadgil has a PhD from the Electrical Engineering Department of IIT Bombay. She works with the LTE-Wi-Fi Interworking group of the EE Department, IIT Bombay and for the project funded by Department of Electronics & Information Technology (DeitY). She is currently the Head of Computer Technology Department, Vartak Polytechnic, Vasai, Maharashtra. She has graduated from Pune university with merit rank. She completed her MTech in Electronics Engineering from VJTI and received the ISTE L&T National Award for the best MTech thesis completed at VJTI and SAMEER, IIT Bombay. She has published papers in Flag-ship National and International conferences. She has text books and laboratory manuals published for the Maharashtra State Board of Technical Education, Mumbai for Microprocessors, Operating Systems and Advanced Microprocessors.

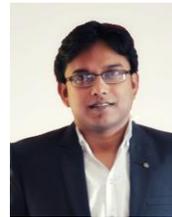

Shashi Ranjan: Shashi Ranjan is currently pursuing his PhD from the Electrical department, IIT Bombay. He completed his MTech from IIT Bombay in 2013 and BTech from BIT Mesra in 2010. During 2013-2015 period, he worked as Project Research Engineer at IIT Bombay on the project Dual Connectivity LTE-WiFi Solution for Broadband Wireless, funded by Department of Electronics & Information Technology (DeitY). He has also actively participated and contributed towards Telecommunications Standards Development Society India (TSDSI) and 3rd Generation Partnership Project (3GPP). He has been IIT Bombay representative at 3GPP RAN2 and SA6 Working Group. His research interests include Next generation networks, Fog computing, Mobile communications and others.

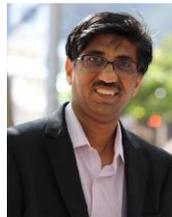

Abhay Karandikar Prof Abhay Karandikar is currently Dean (Faculty Affairs) and Institute Chair Professor in the department of electrical engineering at Indian Institute of Technology (IIT) Bombay. He was the Head of the Department from January 2012 to January 2015. He is the coordinator of Tata Teleservices IIT Bombay Center for Excellence in Telecom (TICET), National Center of Excellence in Technology for Internal Security. He spearheaded a national effort in setting up Telecom Standards Development Society of India (TSDSI), India's standards body for telecom with participation of all stakeholder. Prof Karandikar is the founding member and Chairman of TSDSI. Currently, he is working as a consultant to provide technical expertise in design and implementation of BharatNet, Government of India's flagship initiative under Digital India program. Dr. Karandikar has several patents issued and pending, contributions to IEEE, 3GPP standards, contributed chapters in books and large number of papers in international journals and conferences to his credit. Dr. Karandikar was awarded with IEEE SA's Standards Medallion in December 2016 in New Jersey. His team also won Mozilla Open Innovation challenge prize in March 2017. He is a two time recipient of award for excellence in teaching at IIT Bombay- in 2006 and 2011. He is co-author of papers which won the Best paper awards in ACM MobiHoc 2009, Workshop on Indoor and Outdoor Small Cells WiOpt2014 and finalist for the best paper award in IEEE LCN 2012 and IEEE NCC 2014 conferences.